# Strategic Issues For A Successful E-Commerce

**Debajyoti Mukhopadhyay, Sangeeta Mishra**
Web Intelligence & Distributed Computing Research Lab
Green Tower, C-9/1, Golf Green, Calcutta 700095, India
Email: {debajyoti.mukhopadhyay, sangeeta.mishra}@gmail.com

*E-commerce is an emerging technology. Impact of this new technology is getting clearer with time and results are tangible to the user community. In this paper we have tried to focus some of its issues like paradigms, infrastructure integration, and security, which is considered to be the most important issue in E-Commerce. At first we have elaborated the paradigms of E-Commerce (Business-to-Business and Business-to-Consumer). Then comes the necessity of infrastructure integration with the legacy system. Security concerns comes next. Rest of the part contains conclusion and references.*

## 1. INTRODUCTION

Electronic Commerce is for the age of Information Technology what mercantilism, the quest for gold and the conquest of new lands were for the age of discovery. Like the prow of a large fishing boat, it draws towards itself all other interests and elements of society, and it will leave new discoveries and changes in its wake. The vast networking of world through optic fibers, satellites and wireless communication is creating a new global community and a new global market, in which most of the countries should participate. It is strengthening, almost paradoxically, the identity of small groups, isolated communities and minority interests and driving them towards a less costly social and economic activity and widening their opportunities. And most importantly, it is empowering small businesses to compete with multinational corporations and enabling consumers to search the world for exactly what they needed.

E-Commerce basically means using networks (Internet) to carry out all the activities involved in business management and operation: buying and selling of products and services, technology and partner search, dealing with counterparts, choosing the most convenient transportation and insurances, performing bank transactions, paying and billing, communicating with company salesmen, picking up orders, and any other activities necessary for trading.

A company will be able to post a complete catalog of it's products and services on the Internet, which can be continuously updated to present new or updated products, proving a large virtual showcase for potential clients, a means to communicate with clients and in that way, adjusts it's offer to their requirements; while at the same time it will get access to virtual markets where it can purchase what it needs.

Through integral systems already under development, one company will connect to other companies located anywhere in the world, to buy and sell, choosing the products and services which best meets its needs from a huge network.
And it's true that this revolution involves us all.

## 2. BUSINESS-TO-BUSINESS (B2B)

**B2B** e-commerce means companies buying from and selling to each other online. It automates and streamlines the process of buying and selling the intermediate products. It provides more reliable updating of business data. **B2B** makes product information available globally and updates it in real time. Hence, procuring organization can take advantage of vast amount of product information. [3]

Now, we must know what are the entities of **B2B** e-commerce & their concerns:

- **Selling company:** with marketing management perspective.
- **Buying company:** with procurement management perspective.
- **Electronic intermediary:** A third party intermediating service provider (the scope of service may be extended to include the order fulfillment ).
- **Deliverer:** who should fulfill the **JIT** (Just in Time Delivery)
- **Network platform:** such as the Internet, Intranet, and Extranet.
- **Protocols and communication:** such as **EDI** (Electronic Data Interchange) and comparison shopping, possibly using software agents.
- **Back-end information system:** possibly implemented using the intranet and **E**nterprise **R**esource **P**lanning (**ERP**) systems.

**B2B** e-commerce implies that both the sellers and buyers are business corporations. It covers a broad spectrum of applications that enable an enterprise or business to form electronic relationship with their distributors, re-sellers, suppliers, and other partners. **B2B** applications will offer enterprises access to the following sorts of information :

- **Product:** Specifications, prices, sales history.
- **Customer:** Sales history and forecasts.
- **Supplier:** Product lines and lead times, sales terms and conditions.

- **Product process:** Capacities, commitments, product plans.
- **Transportation:** Carries, lead-times, costs.
- **Inventory:** Inventory levels, carrying costs, locations.
- **Supply chain alliance:** Key contacts, partners roles and responsibilities, schedules.
- **Competitor:** Benchmarking, competitive product offering, market share.
- **Sales and marketing:** Point of sales (POS), promotions.
- **Supply chain process and performance:** Process descriptions, performance measures, quality, delivery time and customer satisfaction.

## 2.1 How to get the best

People always want to get the best shot in life. To deliver a sound return on your investment you must add on time delivery and flavor of some strategies. This strategy should include proper marketing, channel management, solid technology, strategic partners and great products. Let us have a look on each of them.

### 2.1.1 Just in Time delivery (JIT)

In such a case **(JIT)**, delivery materials and parts on time is a must. Using E-Commerce, it is highly possible to assure **JIT** deliveries. Just in time delivery can be realized by the co-coordinated effort of delivery- service company and suppliers inventory policy.

Quick delivery does not necessarily mean **JIT** delivery, but the system for quick delivery is the backbone of **JIT** delivery. For the **B2B** E-Commerce environment the advance confirmation of the delivery date at the contract stage is very important. [5][15]

### 2.1.2 Add strategies to your business

#### 2.1.2.1 Direct Marketing

In a typical business organization, buying decisions, especially for products over a few thousand dollars, are made by group of individuals. As a result, direct marketers need to extent the reach of their programs to different functional areas and perhaps even different levels within a functional area.

There are multiple buyers and influences in any organization who play a role in the buying decision. You may know with reasonable certainty who your primary target is, but secondary target can be just as important to reach. You may have to reach business buyers and influencers in three basic management areas (functional management, financial management and general management) and do it at middle to upper managerial, as well as technical levels. To do it companies need accurate E-mail list, which they can develop by viewing companies Websites and reviewing annual reports and other public documents.

### 2.1.2.2 Relationship Marketing

Business buyers are not always ready to buy products or services when you are ready to sell them. Factors you cannot control, such as the companies' budgeting process, the need for additional approvals, or purchasing procedures, may have a direct impact on plans to purchase. There may be a casual interest in the product but not an immediate need.

The smart **B2B** direct marketer compensates for this uncertainty by making sure a program of regular, ongoing communications (often called a continuity program) is in front of prospects periodically. This can be done by direct E-mail and by placing the information on the website.

### 2.1.2.3 Internet Marketing

Several potential marketing strategies can be used in **B2B** E-Commerce marketing. These strategies can be classified into the following five categories:
1. Generating and qualifying leads with the Internet.
2. Using Internet events to promote products and services.
3. Executing instant fulfillment on the Internet.
4. Generating orders through the Internet.
5. Enhancing customer relationship with the Internet.

### 2.1.2.4 Channel Management

The first element is coherent marketing or channel management. The true test of a successful E-Commerce implementation is how well it exploits the Internet to reach, capture and retain the right customers. Choosing which products and services will be offered through which channel is also a crucial decision.

E-commerce runs across multiple sales channels, including direct, indirect and E-marketplaces. The choice of which marketplaces to use as sales channels is a crucial decision.

In addition to marketplaces, using indirect sales channels is also an area for explosive sales opportunities. Enabling your selling partners to host your catalog, inventory and fulfillment databases on their systems can create efficiency that grows their business and yours. You also can continue your direct one-to-one trading relationship with long time strategic vendors by "E-enabling" the entire business process from the initial request for quote through order fulfillment to automatic billing and payment.

These channels create a situation where the E-Commerce sell side platform must transact across multi-channel selling strategies --which brings us to the next element of your strategy: technology.

### 2.1.2.5 Technology

Industry standard tools often allow a seller to build and manage product catalogs and content once and use them throughout the entire multi-channel selling conduits. Evolving tools and capabilities allow you to develop customer friendly web sites and win repeat customers by building customer loyalty. The front end for e-commerce selling is an important piece of **B2B** success, connecting your new web systems with your existing systems. The 24*7 online marketplace means your E-business has to be continually available. **IT** infrastructure must provide more performance, reliability, security and process integration than a bricks-and-mortar environment. In addition, mainframes hosting the databases and **ERP** (**E**nterprise **R**esource **P**lanning) systems operating the management systems must be seamlessly integrated with the e-commerce engine to provide the caliber of service customers expects and to realize the cost efficiencies **B2B** E-Commerce can provide. Choosing a flexible E-Commerce platforms and a system integrator experienced with the entire business process is a must for success.

### 2.1.2.6 Partners

Like choosing the Internet as a sales channel, it's also important to select the right partners, including an integration partner who is experienced in helping to move ahead rapidly across the entire E-business process. we have to accept that any move to E-Commerce is not about incremental improvement, rather fundamental redesign of the key business processes.

### 2.1.2.7 Products

With the presence on the web, we can effectively and efficiently transact business with our clients 24*7. But so can our competitors. Survivals and success in E-Commerce entails more than simply building a storefront to sell online. [5]

### 3. BUSINESS-TO-CONSUMER (B2C)

While the term E-Commerce refers to all online transactions, **B2C** stands for **"Business-to-Consumer"** and applies to any business or organization that sells its products or services to consumers over the Internet for their own use.

In the late 90s, dotcoms— which were quickly gaining in size and market capitalization — posed a threat to traditional brick and mortar businesses. In many ways, these dotcoms seemed to be rewriting the rules of business — they had the customers without the expenses of maintaining physical stores, little inventory, unlimited access to capital and little concern about actual earnings. The idea was to get big fast and worry about profits later. And a popular thought automatically comes into our mind: " Learn to swim while the tide is out. Learn from the kinds of customers that are out there now. It is a small market- play with it; learn to price business in this market, learn how to assess risk. If you can do it well, the stakes will get higher and you will succeed where others may not."

**3.1 What are the major challenges of B2C e-commerce**

- **Getting browsers to buy things** — Your E-Commerce site cannot live on traffic alone. Getting visitors to the site is only half the battle. Whether they buy something is what determines if you win.

  Some ways to boost the **B2C** conversion rate include improving navigation, simplifying checkout process (such as one-step checkout and easily replaced passwords), and sending out e-mails with special offers.

- **Building customer loyalty** — With so many sites out there, how can you build a strong relationship with customers? Here are some tips:

  - Focus on personalization: A wide array of software packages are available to help e-commerce sites create unique boutiques that target specific customers.

  - Create an easy-to-use customer service application. Providing just an e-mail address can be frustrating to customers with questions. Live chat or, at the very least, a phone number will help.

  - Focus on making your site easy to use.

- **Fulfillment** — E-Commerce has increased the focus on customer satisfaction and delivery fulfillment. Companies should improve their logistical systems in order to guarantee on-time delivery. Providing instant gratification for customers still isn't easy, but successful **B2C** E-Commerce operations are finding that fulfillment headaches can be eased with increased focus and investment in supply chain and logistical technologies. [5]

**3.2 Six Keys to B2C E-Commerce success**

So, what does it really take to capture the E-consumer and generate online insurance sales? Based on Insurance & Technology's interviews with both early adopters and industry analysts, there appear to be six key success factors:

i. **Strategic Goals Assessment/Customer Needs Assessment**

What are your goals as a company? Who are your customers? What are their needs? These may sound like basic questions, but both insurers and analysts emphasize that a company's Web presence must reflect this information.

ii. **Create a Usable, Targeted and Sticky Web Site**

Usability and site performance are some of the key factors insurers need to keep in mind when developing their **B2C** E-Commerce strategies. Insurers also need to be aware of all of their various constituencies when developing **B2C** initiatives. The Web can reach multiple audiences and none should be overlooked. A good Web site will communicate with consumers as well as business partners, agents, suppliers and vendors.

Stickiness, or the success of a Web site in attracting and keeping new and returning visitors, is another success factor. Turning the site into more of an information portal with real-time news feeds with keeping content updated and synchronized will help keep customers coming back.

iii. **Integration**

The Internet is not a stand-alone platform or medium. To be an effective service and distribution channel, it must be integrated with back-end legacy systems, agent systems, call centers, marketing initiatives and pricing and underwriting systems. The Internet is simply another customer relationship channel and integration with other customer service functions is definitely a number-one priority.

iv. **Innovate with Web Applications and Real-Time Transactions**

**B2C** online applications range from the relatively basic, such as updating policy information, to the complex, such as comparative rate quoting and electronic claims submission. Regardless of the specific functions a company plans to add to its Web site, they must serve the needs of the E-consumer. This means that web sites should have interactivity and immediate gratification.

v. **Partnerships**

Although insurers need to be selective in initiating online partnerships, such agreements have the potential to extend market reach and add features in a relatively low-cost manner. According to a recent Gartner Group study, 46 percent of insurance firms active on the Web have partnerships with banks, 30 percent have partnerships with other insurance companies and 22 percent have partnerships with investment firms. Partnerships with insurance portals provide comparative quoting capabilities and may generate business.

vi. **Put Tools in Place To Keep Learning**

E-consumer is a moving target. Investor should always say that they are still playing and always capturing information from all of their channels. They must focus on groups, used third-party assessments and have hired user interface specialists. "The process is iterative: You just keep learning." [5]

## 4. INFRASTRUCTURE INTEGRATION

In this web enabled world, customers rule. The ability to offer mass customization has become a practical reality. To rapidly meet these requirements, time to deployment of new or enhanced application is shrinking dramatically. These applications must be built to be easy to use, nimble, open, extensible, and available across all platforms and all these demanding characteristics must be achieved at minimal cost.

Replacement of legacy system application is costly so it is seen that people started to aggregate information from disparate sources and integrate them for seamless information flow, the demand to communicate with a wide variety of mobile devices, and the shortage of skills and knowledge that are further compounded by shrinking time-to-deployment requirements.

These software integration technologies lower development and deployment costs by doing the following:

- Supplying the communication and integration code so application developers can concentrate on the value-added business logic;
- Providing a standard platform on which to build, deploy, and manage distributed applications;
- Reducing the **IT** skills required to deliver difficult enterprise requirements;
- Providing rapid application development tools to eliminate custom coding and simplify integration; and
- Enabling the reuse of integration components over many projects.

### 4.1 What is needed for Integration

i. **Requirements**

Traditional requirements definition based on the functionality desired is yielding to a definition Based on time-to-deployment and the ability to integrate future technologies. New infrastructure requirements are emerging which places more importance on the task of planning the migration path to newer technologies.

ii. **Technology selection**

For the best result the right technology should always be picked up. Technology should be such that the integrated solution fulfill the following criteria's: Extensibility and reusability, Flexibility, Efficiency, Interoperability and breadth, Cost effectiveness, Ease of maintenance, Deployment ease and efficiency, Ease of administration, Industry acceptability, Enterprise integration, Technological innovation. [1]

iii. **Organizational preparation**

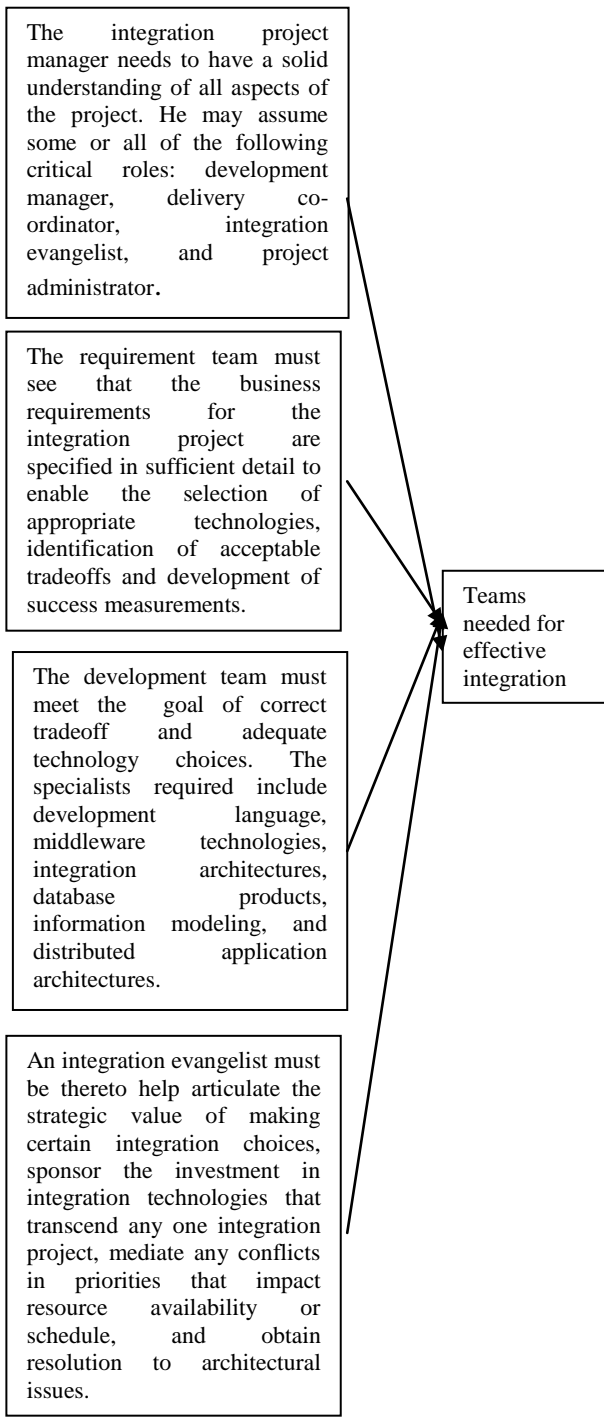

The integration project manager needs to have a solid understanding of all aspects of the project. He may assume some or all of the following critical roles: development manager, delivery co-ordinator, integration evangelist, and project administrator.

The requirement team must see that the business requirements for the integration project are specified in sufficient detail to enable the selection of appropriate technologies, identification of acceptable tradeoffs and development of success measurements.

The development team must meet the goal of correct tradeoff and adequate technology choices. The specialists required include development language, middleware technologies, integration architectures, database products, information modeling, and distributed application architectures.

An integration evangelist must be thereto help articulate the strategic value of making certain integration choices, sponsor the investment in integration technologies that transcend any one integration project, mediate any conflicts in priorities that impact resource availability or schedule, and obtain resolution to architectural issues.

Teams needed for effective integration

**4.2 Benefits of Integration**

It has been surveyed that the end users are benefited in various ways after the completion of Integration project. Benefits thus obtained are :

1. Simple and complete development platform,
2. Platform independence,
3. Network-aware development and run-time platform,
4. Technologically unified intranet, extranet and Internet,
5. Central administration of new software versions,
6. Easy access to enterprise **IT** resources,
7. Rich and highly functional user interface component,
8. Simple and robust security model. [1]

**5. SECURITY ISSUES**

Security is a major issue in developing E-Commerce because this is probably the most important reason people hesitates to buy things on the Net. Buying on the Net requires your credit card number and other personal information. But broadcasting your credit card number through the ether? It sounds pretty dicey. So, it's a challenge for companies to make their site secure and safe so that people can fully rely on them.

**What does security imply**

Whatever the environment, paper or electronic, securing it necessarily implies the prevention of

- Destruction of information and
- Unauthorized availability of information.

**Security issues**

The issues that confront us in relation to securing electronic transaction are therefore:

- Confidentiality
- Integrity
- Availability

- Authenticity/Non-repudiability
- Auditability

**Confidentiability**

Information should be protected from prying eyes of unauthorized internal users, external hackers and from being intercepted during transmission on communication networks by making it unintelligible to the attacker. The content should be transformed in such a way that it is not decipherable by anyone who does not know the transformation information.

**Integrity**

On retrieval or receipt at the other end of a communication network the information should appear exactly as was stored are sent. It should be possible to generate an alert on any modification, addition or deletion to the original content. Integrity also precludes information "replay" i.e., a fresh copy of the data is generated or resent using the authorization features of the earlier authentic message. Suitable mechanisms are required to ensure end-to end message content and copy authentication.

**Availability**

The information that is being stored or transmitted across communication networks should be available whenever required and to whatever extent as desired within pre-established time constraints. Network errors, power outages, operational errors, application software errors, hardware problems and viruses are some of the causes of unavailability of information. The mechanisms for implementation of counter measures to these threats are available but are beyond the scope of end-to-end message security for implementing Electronic Commerce.

**Authenticity**

It should be possible any person or object from masquerading as some other person or object. When a message is received it should therefore be possible to verify it has indeed been sent by the person or object claiming to be the originator. Similarly, it should also be possible to ensure that the message is sent to the person or object for whom it is meant. This implies the need for reliable identification of the originator and recipient of data.

**Non-repudiability**

After sending / authorizing a message, the sender should not be able to, at a later date, deny having done so. Similarly the recipient of a message should not be able to deny receipt at a later date. It should, therefore be possible to bind message acknowledgements with their originations.

**Auditability**

Audit data must be recorded in such a way that all specified confidentiality and integrity requirements are met.[2]

**Security solutions**

- **Cryptography** is the most widely used technique for implementing technology solution for the above mentioned security problems. It comprises encryption -- the process of making information unintelligible to the unauthorized reader and decryption – reserving encryption to make the information readable once again. Conventional cryptography uses a secret code or key to encrypt information. The same secret key is used by the receiver to decrypt the information.[14]
- **Password** is the most common mechanism used for authenticate people. Passwords are expected to be known only by the owner. The onus is on the owner to keep the password secret.
- **Digital signature** can be used not only to verify the authenticity of the message and the claimed identity of the sender, but also to verify the message integrity. The recipient, however, should not be able to use the received digital signature to falsely "sign" messages on behalf of the original sender. Here a message is encrypted with the sender's private key to generate the 'signature'. The message is then sent to the destination along with this signature. The recipient decrypts the signature using the sender's public key, and if the result matches with the copy of the message received, the recipient can be sure that the message was sent by the claimed originator and that the message has not been modified during transmission, since only the originator is in possession of the corresponding encryption key. It is a two key cryptosystems.
- A more effective solution can be obtained by using a **biometric** authentication device, such as a fingerprint scanner, in the e-wallet.
- **Smart card** are similar to credit cards except that they have chips embedded in them. These cards can be used to store value and carry authentication information.

# 6. CONCLUSION

Changing market scenario puts pressure on business persons to adapt new and smart strategies to reach the pinnacle of success. New inventions are rapidly becoming part of IT infrastructure. But to get effective feedback we need a multifunctional team approach consisting of business people who can correctly identify business requirements, technology requirements and success criteria. People can reduce the risk and time-to-deployment by considering the factors described above.

## 7. REFERENCES

1. Aberdeen Group, Inc. "e-Business Infrastructure Integration: Practical Approaches," An Executive White Paper, Boston, Massachusetts 02108, USA, November 2001
2. Kamlesh K. Bajaj and Debjani Nag, "E-Commerce: The Cutting Edge of Business," Tata-McGrawHill, 1999
3. Efraim Turban, Jae Lee, David King, H. Michael Chung, "Electronic Commerce-A Managerial Perspective ," Pearson Education Asia, 2001
4. Ravi Kalakota and Andrew B. Whinston, "Frontiers of Electronic Commerce," Addison Wesley, 2001
5. Susmita Das, Malabika Dinda, Sudipa Batabyal, Sangeeta Mishra, "A Study on Various Aspects of E-Commerce Paradigms with One Design Implementation," B.Tech.(Honours) Thesis, Haldia Institute of Technology (Vidyasagar University), 2002
6. Simon S. Y. Shim, Vishnu S. Pendyala, Meera Sundaram, and Jerry Z. Gao, "Business to Business E-Commerce Frameworks," Computer, IEEE Computer Society, Volume 33, Number 10, October 2000.
7. Wenli Wang, Zoltan Hidvegi, Andrew D. Bailey Jr., and Andrew B. Whinston, "E-Process Design and Assurance Using Model Checking," Computer, IEEE Computer Society, Volume 33, Number 10, October 2000.
8. Tim Ebringer, Peter Thorne, and Yuliang Zheng, "Parasitic Authentication To Protect Your E-Wallet," Computer, IEEE Computer Scciety, Volume 33, Number 10, October 2000.
9. Abhijit Chaudhury, Debasish Mallick, and H. Raghav Rao, "Web Channels in E-Commerce," Communications of the ACM, Volume 44, Number 1, January 2001.
10. Ted Becker, "Rating the Impact of New Technologies on Democracy," Communications of the ACM, Volume 44, Number 1, January 2001.
11. Joe Mohen and Julia Glidden, "The Case for Internet Voting," Communications of the ACM, Volume 44, Number 1, January 2001.
12. Deborah M. philips and hans A. von Spakovsky, "Gauging the Risks of Internet Elections," Communications of the ACM, Volume 44, Number 1, January 2001.
13. Lance J. Hoffman and Lorrie Cranor, Guest Editors, " Internet Voting for Public Officials," Communications of the ACM, Volume 44, Number 1, January 2001.
14. Andrew S. Tanenbaum, "Computer Networks," Third Edition, Prentice-Hall.
15. Debajyoti Mukhopadhyay and Sangeeta Mishra, "How to Meet The Challenges Of Managing E-Commerce Successfully," Journal of the Calcutta Management Association, Volume VII, Number 2: August 2002.